\documentclass[twocolumn,aps,prb,showpacs]{revtex4}
\usepackage{epsfig,graphicx,times}

\usepackage{amssymb}
\usepackage{amsmath}
\usepackage{graphicx}
\usepackage{bm}

\begin{document}
%\draft

\title{Experimental Upper Bound on Superradiance Emission from Mn$_{12}$ Acetate}

\author{ M. Bal$^1$, Jonathan R. Friedman$^{1,*} $, K. Mertes$^{1,2}$, W. Chen,$^3$ E. M. Rumberger$^4$, D. N. Hendrickson$^4$, N. Avraham$^{2,5}$, Y. Myasoedov$^5$, H. Shtrikman$^5$ and E. Zeldov$^5$}
\affiliation{ \mbox{$^1$ Department of Physics, Amherst College, Amherst, Massachusetts 01002-5000} \\
\mbox{$^2$Physics Department, City College of the City University of New York, New York, New York 10031}
\mbox{$^3$Department of Physics and Astronomy, Stony Brook University, Stony Brook, New York 11794-3800}
\mbox{$^4$Department of Chemistry and Biochemistry, University of California at San Diego, La Jolla, California 92093}
\mbox{$^5$Department of Condensed Matter Physics, The Weizmann Institute of Science, Rehovot 76100, Israel}}

\date{\today}
\begin{abstract}

We used a Josephson junction as a radiation detector to look for evidence of the emission of electromagnetic radiation during magnetization avalanches in a crystal assembly of Mn$_{12}$-Acetate. The crystal assembly exhibits avalanches at several magnetic fields in the temperature range from 1.8 to 2.6 K with durations of the order of 1 ms. Although a recent study shows evidence of electromagnetic radiation bursts during these avalanches [J. Tejada, et al., Appl. Phys. Lett. {\bf 84}, 2373 (2004)], we were unable to detect any significant radiation at well-defined frequencies. A control experiment with external radiation pulses allows us to determine that the energy released as radiation during an avalanche is less than 1 part in 10$^4$ of the total energy released. In addition, our avalanche data indicates that the magnetization reversal process does not occur uniformly throughout the sample.

\end{abstract}

\maketitle

Single-molecule magnets have been subject to intense experimental and theoretical investigations during the past decade. At low temperatures, they are bistable and exhibit hysteresis similar to classical magnets.\cite{18, 10} In addition, they exhibit fascinating quantum mechanical properties such as tunneling between "up" and "down" orientations,\cite{33, 81, 91} and interference between tunneling paths.\cite{162} Furthermore, their potential use in quantum computation has been proposed,\cite{298} and experiments with millimeter-wave radiation have been carried out showing that the relaxation rate for magnetization reversal,\cite{285, 283} and the energy-level populations can be controlled.\cite{333, 334, 336} Chudnovsky and Garanin,\cite{335} have proposed theoretically that single-molecule magnets could be used to generate superradiance.   In a recent experiment, Tejada et al. reported that during magnetization avalanches of the molecular magnet Mn$_{12}$ acetate millimeter-wave radiation was released.\cite{332} They interpret the results as evidence of superradiance in the system, in which coherent radiation is produced at frequencies corresponding to transitions between spin states of the molecule.  The bolometer used in that study, however, was not sensitive to radiation frequency and so no information about the spectrum of the radiation was available.  Here we report on a study of the fast dynamics of magnetization reversal during avalanches in a crystal assembly of Mn$_{12}$ acetate.  We used a Josephson junction as a radiation detector, exploiting the AC Josephson effect to create a frequency-sensitive detector.  We were unable to detect any radiation at the predicted superradiance frequencies during magnetization avalanches.  Using external radiation, we were able to set limits of the power and duration of radiation that might be emitted during an avalanche.

Magnetization avalanches in Mn$_{12}$ were found at low temperatures\cite{15} even before resonant magnetization tunneling was discovered in this material.\cite{33} In brief, the avalanches are thermal runaway events in which the heat released by the reversal of a metastable spin in an external field induces other spins to flip, rapidly producing a total saturation of the magnetization.  Since tunneling increases the rate at which spins flip, avalanches tend to occur at the magnetic fields where resonant tunneling occurs.  Avalanches are seen in samples that are thermally well isolated from their environment:  the reduced heat flow out of the sample allows it to reach a temperature high enough to induce an avalanche.  Thus avalanches have been seen at very low temperatures,\cite{15} where heat capacities and thermal conductivities are low, and in large crystal assemblies, in which the sample's surface-to-volume ratio (and hence the thermal conductivity) is small.  

The proposed mechanism of superradiance in Mn$_{12}$ is represented in Fig.~3 of Ref.~\onlinecite{332}.  The avalanche heats up the sample, creating a transient increase in the population of high-lying levels.  This population can decay either by the emission of phonons or of photons.  Normally, the former is favored.  However, under the proposed superradiance mechanism, if the wavelength of the radiation is large compared to the size of the sample, then the excited spins can decay coherently, leading to a substantial enhancement in the radiation emitted, with the radiation power being proportional to $N_m^2$ , where $N_m$ is the number of spins in the radiating state  $\left|m\right\rangle$.  The transitions that favor superradiance are between higher-lying levels, where the associated wavelength is large.  Tejada et al. measured a radiation burst simultaneous with an avalanche.  In accordance with the proposed model, if the radiation is due to superradiance, it should primarily comprise a few frequencies corresponding to transitions between high-lying levels.

Josephson junctions can be used as sensitive detectors of electromagnetic radiation and can provide spectral information. When monochromatic electromagnetic radiation is incident on a Josephson junction, the current-voltage curve exhibits steps, known as Shapiro steps.\cite{1} The steps occur at voltages that are integer multiples of $h \nu /2e$, where $\nu$  is the frequency of the radiation.  The Josephson junction is a non-linear device so that if it is exposed to more than one frequency, it will also show steps at voltages corresponding to the sums and differences of the applied frequencies.  If broadband radiation is applied, the steps will be washed out; however, the junction's critical current will still be suppressed.

Several single crystals of Mn$_{12}$, with a total weight of $\sim$20 mg ($\sim$6 x 10$^{18}$ molecules), were assembled together into an approximate cube of length $\sim$3 mm.  The crystals' anisotropy axes were aligned along the direction of the applied magnetic field to within a few degrees. The magnetization of the crystal assembly was measured by a 50 x 50  $\mu$m$^2$ Hall-bar detector. A Nb/AlOx/Nb Josephson junction was placed several inches away from the sample where the magnetic field strength is roughly one tenth of the field that the sample is experiencing. In order to measure the efficiency of the junction as a radiation detector, externally controlled radiation was applied via a rectangular waveguide.  The distance from the end of the waveguide to the junction was nearly equal to the distance from the sample to the junction.\cite{note}

\begin{figure}[htb]
\centering
\includegraphics[width=80 mm]{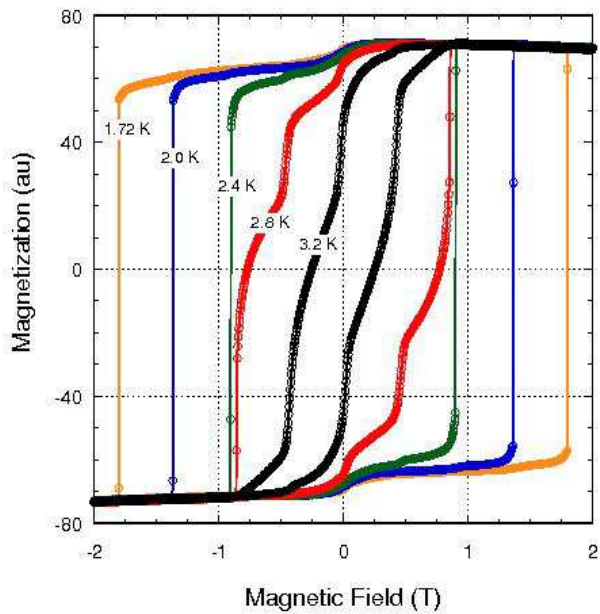}
%\epsfbox{Fig1.eps} %\vskip 20 pt
\caption{(Color online) Selected hysteresis loops for a Mn12 crystal assembly exhibiting avalanches at various temperatures, as indicated. 
\label{fig1} }
\end{figure} 

Hysteresis curves of the crystal assembly were measured at several temperatures, as shown in Fig.~1. Conventional lock-in techniques were used for magnetization measurements while the field was swept at a rate of 5.5 mT/s. The crystal assembly exhibits avalanches at several magnetic fields, which coincide with the fields at which resonant tunneling of magnetization occur, in the temperature range from 1.8 to 2.6 K. The temperature, which is measured with a thermistor mounted near the crystal assembly, rises rapidly up to $\sim$3.7 K during these avalanches, in agreement with previous results.\cite{332} 

We use a triangular waveform to current bias the Josephson junction. The frequency of the current drive is set between 2 to 5 kHz with an amplitude of $\sim$1 mA. We characterized our junction under identical conditions as those in which we observe avalanches. For example, at a temperature of 1.9 K, the junction's V-I curve was measured with an applied field of 1.73 T (at the sample location). A single 0.2 ms pulse of electromagnetic radiation at a frequency of 82.52 GHz is applied while the junction voltage is measured with a fast digitizing oscilloscope.  This frequency is close to one of the predicted superradiance frequencies.  The data are shown in Figure 2.  Because of the triangular current bias, each leg of the quasi-triangular waveform in the figure represents a V-I curve.  In part (a) of the figure, where the external radiation power was 1.259 mW (measured at the end of the waveguide), during the radiation pulse the induced Shapiro steps (marked with arrows) are clearly visible in the voltage drop across the Josephson junction. An abrupt spike in the junction voltage marks the turn-on time of the pulse. In addition, the critical current is markedly suppressed during irradiation. In part (b), the radiation power was reduced to 0.346 mW.  Here some of the Shapiro steps are just barely visible, but the critical current is still noticeably suppressed.  Our junctions can detect pulses as short as 20 $\mu$s, but a pulse of at least 50 $\mu$s is needed to extract frequency information.

\begin{figure}[htb]
\centering
\includegraphics[width=80mm]{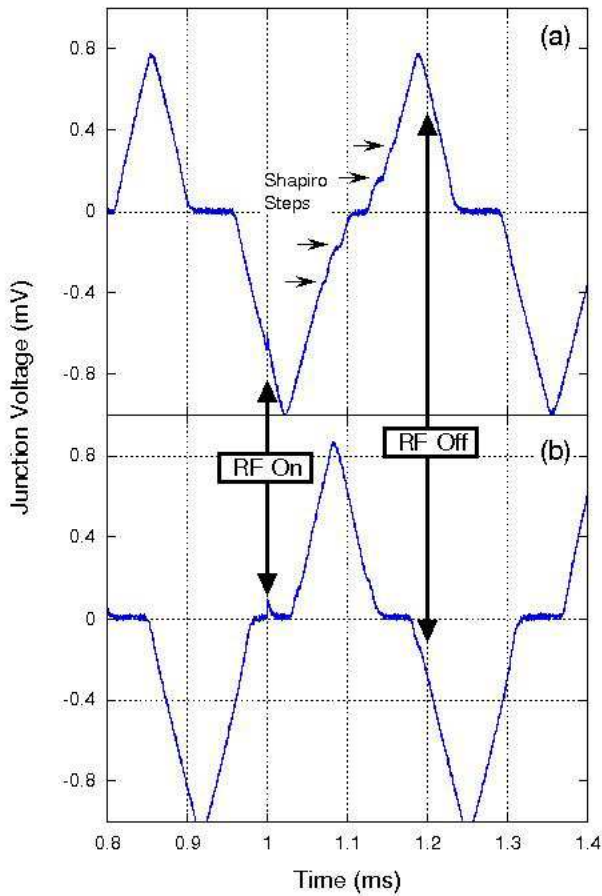} %\vskip 20 pt
\caption{Voltage drop as a function of time across a Josephson junction while a single millimeter-wave 0.2 ms pulse at 82.52 GHz is applied. The junction is current biased with a triangular waveform. The temperature and the magnetic field at the Mn12 crystal assembly location are kept at 1.9 K and 1.73 T, respectively. The radiation power is 1.259 mW in (a) and 0.346 mW in (b).  The relative horizontal displacement between the data sets in (a) and (b) is an artifact produced by the triggering of the oscilloscope. 
\label{fig2} }
\end{figure} 

We have found that the magnetization reversal during avalanches occur on millisecond time scales. Therefore, we used a differential amplifier with a settling time of a few $\mu$s to measure the magnetization avalanches.  The Hall bar was driven with a dc current and a fast digitizing oscilloscope was triggered by the magnetization avalanche signal, allowing us to simultaneously capture the magnetization of the crystal assembly and the Josephson junction voltage during an avalanche, as shown in Figure 3a. The duration of these avalanches increases with the temperature. For instance, at 2.5 K the magnetization switches in 3.7 ms, whereas it takes only 1 ms at 1.9 K. These switching times for Mn$_{12}$ crystal assembly are significantly faster than the 0.1 s reported in Ref.~\onlinecite{332}.  (More recent experiments by the same group give avalanche times consistent with our results.\cite{338})

Despite numerous measurements on avalanches at different temperatures (and therefore different fields - Fig.~1), we did not observe any indications of a radiation burst in the Josephson junction voltage. In fact, we cannot detect any suppression of the junction's critical current during the avalanche, as would be expected with broadband radiation.  Following Ref.~\onlinecite{332}, we can estimate the amount of radiation power expected from just one superradiance transition, the m = 1 to 2 transition, to be $P = N_1 h\nu \Gamma _{SR} $, where $N_1  = N\exp \left( { - U_{eff} /T} \right)$ is the number of spins in state m = 1 (given a total of $N$ spins) and  $\Gamma _{SR} = N_1 \Gamma _1 $ is the superradiance decay rate, with $\Gamma _1 $ being the radiative spontaneous decay rate of the m = 1 state. Using an effective activation barrier $U_{eff} $ of 46 K, $N \sim$ 6 x 10$^{18}$, $T$ = 3.7 K and  $\Gamma _1 \sim$ 1 x 10$^{-7}$ s$^{-1}$, we obtain $P \sim$3 mW, which should be easily detected by our junction.   Superradiance is expected to last about as long as the avalanche rather than $1/\Gamma _{SR}$  because heating from the avalanche repopulates the radiating level.  From our experiment we can estimate an upper limit for the fraction of energy that would be emitted as radiation due to superradiance. The total energy released ($H \cdot \Delta M $) for the crystal assembly during the avalanche at 1.9 K and 1.73 T is $\sim$3.7 mJ.  This translates into an average power of $\sim$3.7 W during an avalanche with a typical duration of $\sim$1 ms. Considering that our Josephson junctions are sensitive to radiation power levels of at least 0.346 mW, we conclude that less than 1 part in 10$^4$ of the total energy released is emitted as photons due to superradiance. The power levels detected during an avalanche by the InSb bolometer used in Ref.~\onlinecite{332} were less than 1 $\mu$W.\cite{338} 

The magnetization reversal process during an avalanche does not appear to occur uniformly throughout the sample.  When the magnetization is measured from the edge of the crystal assembly, the measured magnetization changes in a mostly monotonic fashion, although there is noticeable structure (Figure 3a). However, when the sample is positioned on top of the Hall bar, the signal shows an oscillatory structure that is completely reproducible (Figure 3b).  This indicates that different regions of the sample (perhaps individual crystals) reverse their magnetization sequentially.  In fact, similar results have been obtained in the reversal of single Mn$_{12}$ crystals in the absence of avalanches.\cite{337}  The non-monotonic behavior in Fig.~3b is due to the fact that regions on one side of the detector produce a Hall voltage of opposite sign to that produced by regions on the other side so that when a region of the sample reverses, it can either increase or decrease the Hall signal, depending on its position.  Thus, it appears that if superradiance is occurring in Mn$_{12}$, it does not involve the coherent reversal of the whole assembly.  Since the superradiance power is proportional to $N^2$, the fact that magnetization reversal occurs sequentially throughout the sample, reducing the effective value of $N$, may drastically reduce the amount of radiation power emitted.  This would explain why we do not see the predicted $\sim$3 mW of power and why the power found by Tejada et al. was so small.  

\begin{figure}[htb]
\centering
\includegraphics[width=80mm]{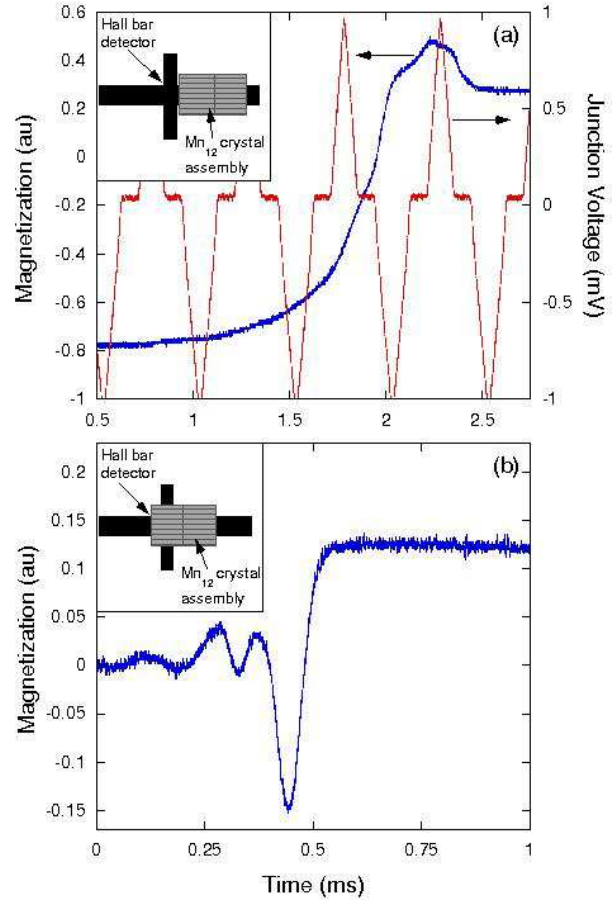} %\vskip 20 pt
\caption{(Color online) (a)  Simultaneously measured magnetization reversal and Josephson junction voltage as a function time during an avalanche at 2.1 K and at the n = 3 tunneling resonance. (b) Magnetization reversal as a function of time during an avalanche at 1.9 K. The relative positions of the crystal assembly and Hall bar are depicted schematically in the insets.
\label{fig3} }
\end{figure} 

Another reason that the emitted power would be suppressed is that superradiance requires that all of the spins involved have the same transition frequencies to within a natural linewidth.  However, because of dipole interactions between molecules, each spin has a different local magnetic field, which in turn leads to a distribution of transition frequencies.  This may sharply reduce the number of spins located within a wavelength of one another that also have the same transition frequency.  So, even within a region of the sample that undergoes uniform reversal the effective number of spins participating in superradiance may be much smaller than the number occupying a radiating level.  

Our results do not rule out the possibility of that superradiance takes place in Mn$_{12}$ avalanches.  However, if it is occurring, less than 10$^{-4}$ of the total energy released goes into radiation at the predicted superradiance frequencies or the radiation pulse is faster than $\sim$50 $\mu$s.  We have found that these avalanches take place in  $\sim$1 ms, much faster than previously reported\cite{332}  and the avalanches show a structure indicative of non-uniform magnetization reversal in the crystal assembly.

We thank M. P. Sarachik, E. Chudnovsky and J. Tejada for useful conversations.  J. Lukens kindly allowed us to use some of his laboratory facilities.  D. Krause and P. Grant made important technical contributions to this study.  Support for this work was provided by the National Science Foundation, the Research Corporation, the Alfred P. Sloan Foundation, and the Center of Excellence of the Israel Science Foundation.


\begin{thebibliography}{10}
\bibitem{note} Radiation from magnetic-dipole transitions should be preferentially emitted in the x-y plane, whereas the junction lies along the z axis from the sample.  Nevertheless, we expect a substantial amount of the emitted radiation to reach the junction since both sample and junction are surrounded by a 1-inch-diameter copper cylinder that lines the inside of the sample chamber and therefore acts as an overmoded waveguide for the radiation.
\end{thebibliography}
\end{document}